\documentclass{jpsj2}
\usepackage{graphicx}
\usepackage{amsmath}

\def\runtitle{%
Influence of impurity-scattering on tunneling conductance  
in ...}  
\def\runauthor {%
N. {\sc Kitaura}, H. {\sc Itoh}, Y. {\sc Asano}, Y. {\sc Tanaka}, J. {\sc Inoue}Y.{\sc Tanuma} and S.{\sc Kashiwaya}
}

\title
{%
Influence of impurity-scattering on tunneling conductance  
in d-wave superconductors with broken time reversal symmetry}

\author{%
Naoki {\sc Kitaura}$^{1}$
Hiroyoshi {\sc Itoh}$^{2}$,
Yasuhiro {\sc Asano}$^{3}$,
Yukio {\sc Tanaka}$^{1,4}$,
Jun-ichiro {\sc Inoue}$^{1}$, 
Yasunari {\sc Tanuma}$^{5}$ and 
Satoshi {\sc Kashiwaya}$^{6}$
}

\inst
{
$^{1}$ Department of Applied Physics, Nagoya 
University, Nagoya, 464-8603, Japan.
\\
$^{2}$ Department of Quantum Engineering, 
Nagoya University, Nagoya, 464-8603, Japan.
\\
$^{3}$ Department of Applied Physics, Hokkaido University, 
Sappro 060-8628, Japan.
\\
$^{4}$ Crest, Japan Science and Technology Corporation (JST), 
Nagoya 464-8063 Japan.
\\
$^{5}$
Institute of Physics,
Kanagawa University,
Rokkakubashi, Yokohama, 221-8686, Japan. 
\\
$^{6}$ National Institute of Advanced Industrial Science and Technology, 
Tsukuba, 305-8568, Japan}

\recdate
{\today}

\abst{%
Effects of impurity scattering on tunneling conductance in
dirty normal-metal/insulator/superconductor junctions are 
studied based on the Kubo formula and the recursive Green function
method. The zero-bias 
conductance peak (ZBCP) is a consequence of the unconventional
pairing symmetry in superconductors. 
The impurity scattering in normal
metals suppresses the amplitude of the ZBCP. 
The degree of the suppression agrees well with 
results of the quasiclassical Green function theory.
When superconductors have $d+is$-wave pairing symmetry, 
the time-reversal symmetry is broken in superconductors and 
the ZBCP splits into two peaks.
The random impurity scattering reduces the height of the two splitting peaks.
The position of the splitting peaks, however, almost remains unchanged even in 
the presence of the strong impurity scattering. 
Thus the two splitting peaks never merge into a single ZBCP. 
 }
\kword{%
Andreev reflection,  $d$-wave superconductor, 
zero-energy states, zero-bias conductance peak,
disorder, $d+is$-wave state 
}
\begin{document}
\maketitle
\section{Introduction}
Charge transport in unconventional superconductors 
shows remarkable properties which cannot be expected 
in conventional $s$-wave superconductors. 
One of the striking effects is the zero-bias conductance peak 
(ZBCP) in normal-metal/unconventional superconductor junctions\cite{Tana1}.
In hybrid structures consisting of high-T$_{C}$ superconductors, 
a number of experiments observed the 
ZBCP~\cite{Kashi95,Kashi96,Alff,Wei,Wang,Iguchi,Kashi2000,Sawa1,Sawa2}.
The zero-energy state (ZES)~\cite{Hu} formed at a surface of the unconventional
superconductor is responsible for
the ZBCP~\cite{Tana1}.
The interference effect of a quasiparticle 
in the presence of the sign-change in the pair potential is the origin
of the ZES~\cite{matsumoto,nagato,ohhashi,sign1,sign2,sign3,Zhu2}. 
The ZES is also responsible for the low-temperature anomaly of the Josephson
current in unconventional superconductor 
junctions~\cite{tanaka6,barash,t61,Kusakabe,asano01-3,asano02-3,asano03-1} and 
anomalies of the charge transport in ferromagnet/unconventional superconductor 
junctions~\cite{Zhu,Kashi2,Zutic,Yoshida,Hirai,Yoshida2002}. 
%

Since the ZES is a result of the interference effect 
of a quasiparticle, it is sensitive to the time-reversal symmetry (TRS)
of systems.
Actually it has been 
confirmed in normal-metal/$d$-wave superconductor junctions that 
the ZBCP splits into two peaks by applying magnetic fields~\cite{fogelstrom,YT021,YT022,YT023}.
It has been also pointed out that the ZBCP 
splits into two peaks
when the TRS is broken in 
superconductors~\cite{fogelstrom,Kashiwaya95,TJ1,TJ2,Tanuma2001,Zhu1}.
Theories showed a possibility of $d+is$-~\cite{matsumoto2} or 
$d+id$~\cite{laughlin}-wave broken time-reversal 
symmetry states (BTRSS) at surfaces of superconductors. 
If such BTRSS's are stable at the surface of superconductors,
they may bring significant influences  
on transport properties~\cite{tanaka3,tanaka4}. 
Experimental results, however, are still controversial.
Some experiments reported the split of
the ZBCP in the absence of magnetic field~\cite{covington,biswas,dagan,sharoni},others did not show the zero-field splitting even in low 
temperatures~\cite{Ekin,Alff,Iguchi,Wei,Sawa1,Aubin}. 
Moreover a experiment of the Josephson junctions concluded the absence of 
the BTRSS at interfaces of $d$-wave junctions~\cite{Neils}. 
Thus it is very important to theoretically address the origin of the splitting. 
There are several factors which determine the magnitude of
splitting of the ZBCP: the transmission probability of the junctions 
\cite{Tanuma2001}, 
the roughness of the interface~\cite{asano02-1}, the impurity scattering in 
superconductors~\cite{asano03-2} and the impurity scattering in normal metals.
Among them, effects of the impurity scattering in normal metals has not been studied yet. 
We note that effects of random potentials on the split of the ZBCP are also
unclear within theoretical studies at present. 
Some theories based on the quasiclassical Green function
method~\cite{eilenberger,larkin,zaitsev,shelankov} did not show
the splitting~\cite{bruder,barash2,golubov,poenicke,yamada,luck}, whereas 
others concluded the 
split of the ZBCP by a numerical simulation~\cite{asano02-1} 
and an analytic calculation~\cite{asano03-2}.
When the BTRSS is formed at the NS interface, a group of experiment
predicted that the two splitting peaks may merge into a single ZBCP
by the impurity scattering~\cite{Greene}. 
\par

%
In dirty normal-metal/insulator/superconductor 
(DN/I/S) junctions, there are many theoretical studies
(see a review~\cite{Beenakker1}) on conventional 
$s$-wave superconductor junctions such as 
numerical methods \cite{Lambert,Takane,Lesovik}, the random matrix theory 
\cite{Beenakker1,Beenakker2,Beenakker3} and the quasiclassical Green function 
approach~\cite{Volkov,Yip,Golubov2,Takayanagi,Nazarov1,Nazarov2,Golubov2003}. 
These studies show a wide variety in line shapes of the conductance depending on
the Thouless energy and the resistance of DN. 
In particular, the ZBCP is an important conclusion in the low transparent junctions even
in the $s$-wave junctions.
We note that the interference of a quasiparticle
while traveling the diffusive metals causes the ZBCP in the $s$-wave junctions.
Recently, the full resistance ($R$) of the disordered 
junctions is derived from a microscopic theory~\cite{Tanaka2002}. 
By applying this theory to $d$-wave junctions, we
show various transport properties which reflect the unconventional 
pairing symmetry in superconductors.
When the $a$ axis of high-$T_c$ superconductors is oriented 45 degrees from
the interface normal, 
$R$ at the zero temperature 
can be expressed as $R_{D} + R_{C}$~\cite{Tanaka2002}, 
where $R_D$ is the resistance of a diffusive metal and $R_C$ is the 
resistance of a clean junction.
The resistance of the disordered junction is given by the simple
summation of the two resistance,
which indicates the absence of the proximity 
effect~\cite{asano01-1,asano01-2,asano02-2}. 
At the present stage, however, 
there is no corresponding studies in DN/I/$d$-wave superconductor 
junctions with the BTRSS. 

In this paper, effects of the 
randomness on the ZBCP in DN/I/S junctions are studied
numerically by using the recursive Green function method. 
We mainly consider the $d+is$-wave pairing symmetry in superconductors. 
For comparison, we also discuss the conductance in the $s$- and $d$-wave junctions 
which are the two limitting cases of the $d+is$-wave junctions.
First we confirm the proximity effect 
in $s$-wave junctions.
The total resistance at the zero-bias in disordered $s$-wave junctions 
shows the reentrant behavior as a function of $R_{D}$, which is consistent 
with the quasiclassical theory.
We also confirm that the resistance in disordered $d$-wave junctions increases
monotonically with the increase of $R_D$~\cite{Itoh1,Itoh2} since there is no
proximity effect. The calculated results are 
consistent with the quasiclassical theory~\cite{Tanaka2002}. 
In $d+is$-wave junctions, we show that the proximity effect appears
when $s$-wave component becomes dominant.
In $d+is$ junctions, the ZBCP splits into two peaks because of the 
BTRSS and the impurity scattering reduces the height of the splitting 
peaks. The two splitting peaks, however, do not merge into a single peak. 
In the light of our theory, the single sharp ZBCP observed in
experiments~\cite{Alff,Wei,Iguchi} is a strong 
evidence for the formation of the pure $d$-wave symmetry state at the interface.  
\par
This paper is organized as follows. 
In Sec. II, the model and method are presented. 
The numerical results are shown in Sec. III. 
In Sec. IV, we summarize this paper.

\section{Model and Method}
\begin{figure}
\begin{center}
\includegraphics[width=8.0cm]{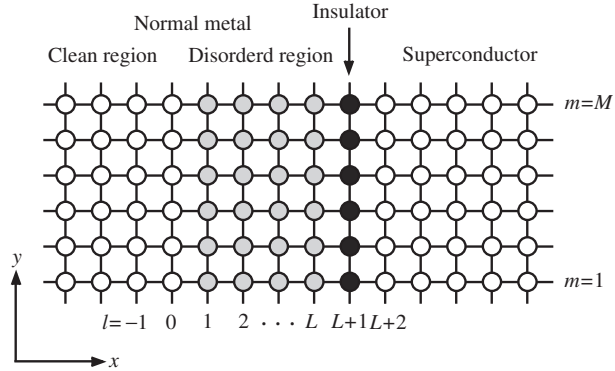}
\end{center}
\caption{
The normal-metal / superconductor junction is illustrated,
where $l$ and $m$ are the lattice indices in the $x$ and the $y$ direction,
respectively. The width of the junction is $M$. We introduce the disordered
region with length $L$ in a normal metal near the interface ( gray circles) and the insulating
barrier at $l=L+1$ ( filled circles). 
}
\label{system}
\end{figure}

Let us consider a normal-metal / superconductor junction on the two-dimensional
tight-binding model as shown in 
Fig.~\ref{system}, where $\boldsymbol{r}=(l,m)$ labels a lattice site. 
We assume the periodic boundary condition in the $y$ direction and 
the width of the junction is $M a_0$, where $a_0$ is the lattice constant. 
The BCS Hamiltonian of the system is expressed as,
\begin{align}
H_{BCS} =& -t \sum_{\langle \boldsymbol{r}, \boldsymbol{r}'\rangle, \sigma}  
\left[ c_{\boldsymbol{r},\sigma}^\dagger c_{\boldsymbol{r}',\sigma}^{ } + H.c. \right] \nonumber \\
&+\sum_{\boldsymbol{r}, \sigma}
(v_{\boldsymbol{r}} -\mu) c_{\boldsymbol{r},\sigma}^\dagger c_{\boldsymbol{r},\sigma}^{ }
 \nonumber \\
&-\sum_{\boldsymbol{r}}\sum_{\boldsymbol{r}'}\left[ 
{\Delta}^\ast(\boldsymbol{r}',\boldsymbol{r})
 c_{\boldsymbol{r}',\uparrow} c_{\boldsymbol{r},\downarrow}
+H.c.\right],\label{bcs}
\end{align}
where $c_{\boldsymbol{r},\sigma}^\dagger$ ($c_{\boldsymbol{r},\sigma}$) is the 
creation (annihilation) operator of an electron at $(l,m)$ with spin 
$\sigma (=\uparrow$ or $\downarrow)$ and $\mu$ is the Fermi energy. 
The summation $\sum_{\langle \boldsymbol{r}, \boldsymbol{r}'\rangle}$ runs over
nearest neighbor sites and $t$ is the nearest neighbor 
hopping-integral.
We assume that the on-site potential $v_{\boldsymbol{r}}$ is zero far from the 
interface in the normal metal and in the superconductor, 
(i.e., $v_{l,m}=0$ for $l\leq 0$ and $L+2\leq l$). 
In the disordered region, $1 \leq l \leq L$, the potential is given 
by random number uniformly distributed in the range of 
\begin{equation}
-\frac{W}{2} \leq v_{l,m} \leq \frac{W}{2}.
\end{equation}
The insulating barrier is described by the potential $v_{L+1,m}=V_{ins}$ for all $m$.
When a superconductor has the $s$-wave pairing symmetry, 
the pair potential is given by
\begin{equation}
\Delta^s(\boldsymbol{r}',\boldsymbol{r})
=\left\{ \begin{array}{ccc} \Delta_s & : & \boldsymbol{r}'=\boldsymbol{r} \\
                           0 & : & \textrm{otherwise}
\end{array}\right..
\end{equation}
When the $a$ axis of a high-$T_c$ superconductor is oriented by 45 degrees from the interface
normal, the pair potential is given by
\begin{equation}
\Delta^d(\boldsymbol{r}',\boldsymbol{r})
=\left\{ \begin{array}{ccc} \Delta_d & : & l=l'\pm 1, m=m'\pm 1 \\
                           -\Delta_d & : & \l=l' \pm 1, m =m' \mp 1 \\ 
                            0        & : & \textrm{otherwise}
\end{array}\right. .
\end{equation}
We note that 
the tight-binding lattices do not 
correspond to the two-dimensional CuO$_2$ plane in high-$T_c$
superconductors. The tight-binding model represents the 
two-dimensional space. 
In our model, we introduce the pair potential between the next nearest 
neighbor sites to describe junctions under consideration. 
An alternative way  
is keeping the pair potential between the nearest neighbor sites and
rotating the square lattice by 45 degrees.
There are no essential differences between results in the two models 
when we focus on the formation of the ZES. 
This is because the ZES is a consequence
of the $d$-wave symmetry of the pair potential.
The BTRSS is characterized by the $d+is$-wave symmetry and the 
pair potential is described by
\begin{equation}
\Delta^{d+is}(\boldsymbol{r}',\boldsymbol{r})
= \Delta^{d}(\boldsymbol{r}',\boldsymbol{r}) \cos\alpha + i 
\Delta^{s}(\boldsymbol{r}',\boldsymbol{r}) \sin\alpha, \label{dis1}
\end{equation}
where $0\leq \alpha \leq \pi/2$ is a parameter which characterizes 
the degree of the time-reversal symmetry breaking. 
When $s$-wave component appears only at a surface of a superconductor,
we describe the pair potential as
\begin{align}
\Delta^{d+is}(\boldsymbol{r}',\boldsymbol{r})
=& \Delta^{d}(\boldsymbol{r}',\boldsymbol{r}) \tanh \left( \frac{x}{\xi} \right) \nonumber \\
& + i 
\Delta^{s}(\boldsymbol{r}',\boldsymbol{r}) \left\{ 1- \tanh \left( \frac{x}{\xi}\right)\right\},
\label{sufs}
\end{align}
where $\xi$ is comparable to the coherence length of $d$-wave 
superconductors \cite{Tanuma2001,matsumoto2}.
In a normal metal, the pair potential is taken to be zero.

The BCS Hamiltonian in Eq.~(\ref{bcs}) can be diagonalized by applying the
Bogoliubov transformation,
\begin{equation}
\left[ 
\begin{array}{c}
c_{\boldsymbol{r},\uparrow} \\
c^\dagger_{\boldsymbol{r},\downarrow} 
\end{array}
\right]= \sum_\nu
\left[
\begin{array}{cc}
u_\nu(\boldsymbol{r}) & -v_\nu^\ast(\boldsymbol{r}) \\
v_\nu(\boldsymbol{r}) & u_\nu^\ast(\boldsymbol{r}) 
\end{array}\right]
\left[\begin{array}{c}
{\gamma}_{\nu, \uparrow} \\
{\gamma}^\dagger_{\nu, \downarrow} 
\end{array}
\right],
\end{equation}
where ${\gamma}^\dagger_{\nu,\sigma}$ 
(${\gamma}_{\nu,\sigma}$) is creation (annihilation) operator of a
Bogoliubov quasiparticle. We omit the spin index of the wavefunction 
($u_\nu,v_\nu$) since we do not consider the spin-dependent potential.
In this way, the Bogoliubov-de Gennes (BdG) equation is derived
on the tight-binding lattice. We solve the BdG equation by using the
recursive Green function method~\cite{lee,asano01-1}.
The Green function is defined in a $2M\times 2M$ matrix form
\begin{equation}
\hat{G}(l,l') = \sum_\nu \frac{ \Psi_\nu(l) \Psi_\nu^\dagger(l')}{E+i\delta -E_\nu},
\end{equation}
where $E_\nu$ is the eigen value of the BdG equation,
$\Psi_\nu(l)$ is the vector with $2M$ components and $m$-th ($m+M$-th)
component is $u_\nu(l,m)$ ($v_\nu(l,m)$).
The differential conductance of the junction is calculated to be
\begin{align}
G_{NS}(eV) =& \frac{e^2}{h} t^2 \textrm{Tr}'\left[ 
\hat{G}(l,l+1) \hat{G}(l,l+1) \right. \nonumber\\ &+ \hat{G}(l+1,l)\hat{G}(l+1,l) 
- \hat{G}(l,l) \hat{G}(l+1,l+1) \nonumber\\ &- \left.\hat{G}(l+1,l+1)\hat{G}(l,l)
\right],
\end{align}
with $E=eV$, where $V$ is the bais voltage applied to the NS junction~\cite{Itoh1,kubo}.
When $\hat{A}$ is a $2M\times 2M$ matrix, $\textrm{Tr}'[ \hat{A} ]$ means 
the summation of $\sum_{m=1}^{M} \hat{A}_{m,m}$.
The conductance in this method is identical to that in the well known 
conductance formula~\cite{blonder,takane92}. 
By averaging over different configuration of randomness, the mean value of 
conductance $<G_{NS}>$ is obtained. 
The conductance in normal conductors $G_{N}$
can be also calculated in the conventional recursive Green function method~\cite{lee}.
The advantage of the method is wide applicability to various
systems such as, the clean (ballistic) junctions,
the dirty (diffusive) junctions and the junctions in the
localization regime~\cite{asano02-4}.

\section{Results}
\begin{figure}
\begin{center}
\includegraphics[width=8.0cm]{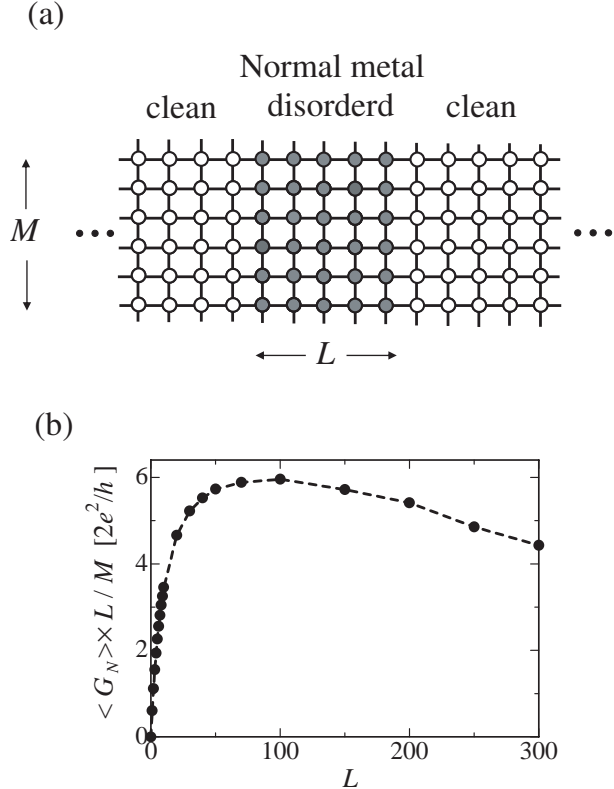}
\end{center}
\caption{
In (a), the system used in a simulation is schematically depicted.
The normal conductance is plotted as a function of the 
length of the disordered region in (b), where $W=2.0t$, $\mu=-1.0t$ and $M=32$.
When the disordered region is in the diffusive regime, the vertical
axis in (b) corresponds to the conductivity.
}
\label{fig:mfp}
\end{figure}

Throughout this paper, we fix the Fermi energy at $\mu=-1.0t$, the amplitude of the
pair potential in the $d$-wave component at $\Delta_d=0.001t$ and
the impurity potential at $W=2.0t$. In order to determine the mean
free path  at the Fermi energy, we first calculate the 
normal conductance as a function of the length of the disordered 
region $L$ in Fig.~\ref{fig:mfp}(b). The ensemble average is carried out 
over 500 samples with different random impurity configurations.
When $L$ is small, $\langle G_N \rangle \sim (2e^2/h) M$. Thus the disordered
region is in the quasiballistic transport regime and 
$\langle G_N \rangle L/M$ is proportional to $L$ as shown in $L<50$. 
In the diffusive regime, $\langle G_N \rangle L/M$ coincides with the 
conductivity which is independent of sample size. Thus $60<L<150$ in
Fig~\ref{fig:mfp}(b) corresponds to the diffusive regime and the mean
free path is estimated to be about 6 $a_0$.
For large $L$, the disordered region is in the 
localization transport regime and the conductance decreases with $L$ in proportional
to $\exp(-L/\xi_{AL})$, where $\xi_{AL}$ is the localization length
of an electron. 

\begin{figure}
\begin{center}
\includegraphics[width=8.0cm]{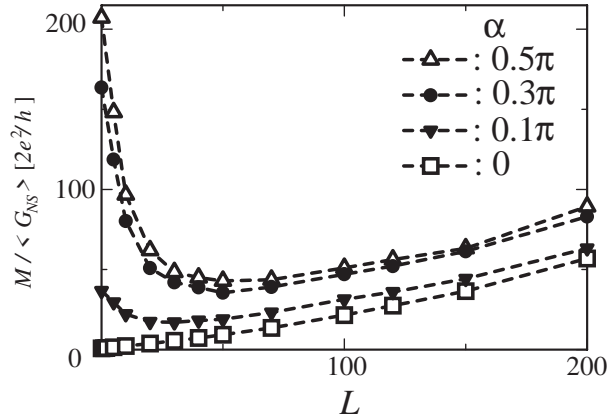}
\end{center}
\caption{
The resistance of the DN/I/S junctions is shown as a function of the
resistance in DN ($R_D$),  where $V_{ins}=5t$ and $M=32$. The pair potential in superconductors 
is given in Eq.~(\ref{dis1}). 
}
\label{fig:pe}
\end{figure}

In Fig.~\ref{fig:pe}, we show the resistance at the zero-bias 
in DN/I/S junctions as a function of $L$.
We assume $d+is$ pairing symmetry in superconductors as shown in 
Eq.~(\ref{dis1}), where $\Delta_s=\Delta_d$.
The results for $\alpha=\pi/2$ correspond to the resistance 
in $s$-wave junctions. The resistance for $\alpha=\pi/2$ 
first decreases with the increase
of $L$. 
Cooper pairs penetrate into normal metals and have a finite amplitude.
This is a source of the proximity effect. As a consequence, 
the reflectionless tunneling suppresses the resistance at the zero-bias. 
For large $L$, the resistance increases linearly with $L$.
The reentrant behavior of the resistance for $\alpha=\pi/2$ is 
the direct consequence of the proximity effect.
The results for $\alpha=0$ correspond to the resistance
in $d$-wave functions. 
The resistance does not show the reentrant behavior and increases almost 
linearly with the increase of $L$. In this case, the proximity effect is absent. 
The amplitude of the Cooper pairs becomes almost zero because
the contribution of the pair potential with positive sign and that with 
negative sign cancel out with each other.
For $\alpha=0.1\pi$, the proximity effect appears because of the $s$-wave
component in the pair potential and the resistance show the weak reentrant
behavior as a function of $L$. 
For $\alpha=0.3\pi$, the reentrant behavior becomes rather remarkable 
and the results are close to those in the $s$-wave junctions.
According to the quasiclassical Green function theory, 
the resistance of the $d$-wave junctions
can be simply given by
\begin{equation}
R=R_{D} + R_{C},
\end{equation}
where $R_{D}$ is the resistance in DN and $R_{C}$ is the resistance
of the clean junction. The summation of these two resistance 
gives the full resistance of the disordered junctions because
there is no proximity effect. The absence of the proximity effect is 
responsible for the anomaly in the Josephson current through 
diffusive metals~\cite{asano01-2,asano02-2}.

\begin{figure}
\begin{center}
\includegraphics[width=8.0cm]{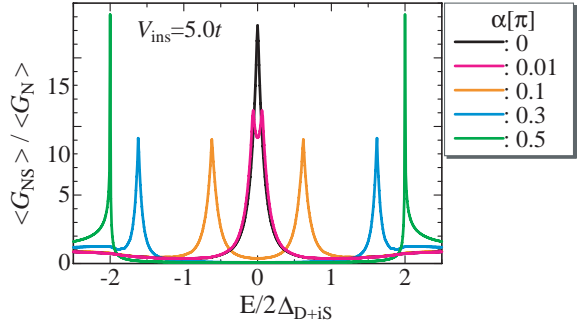}
\end{center}
\caption{
The conductance in clean normal-metal / superconductor junction is 
plotted as a function of the bias voltage,
where $V_{ins}=5t$. The $s$- and $d$-wave junctions are correspond to
$\alpha=\pi/2$ and 0, respectively.
}
\label{fig:dis2}
\end{figure}

In Fig.~\ref{fig:dis2}, we show the conductance in clean N/I/S
junctions as a function of the applied bias voltage, where $V_{ins}=5t$ and $\Delta_s=\Delta_d$.
For $\alpha=0.5\pi$, junctions become $s$-wave junctions and the 
conductance has two peaks at $E=\pm\Delta_s$ which reflect the singularity
of the density of states in $s$-wave superconductors. Junctions with
$\alpha=0$ correspond to $d$-wave junctions and the results show the 
ZBCP owing to the ZES at the junction interface.
For $\alpha=0.01\pi$, the ZBCP splits into two peaks because 
the TRS is broken by the $s$-wave component in the pair potential~\cite{fogelstrom}.
The degree of the splitting increases with increasing $\alpha$.
The calculated results basically remain unchanged when we
assume the $s$-wave component only at the interface.
Since the splitting was observed in several experiments~\cite{covington,biswas,dagan,sharoni}, 
the zero-field splitting of the ZBCP has been believed to be an evidence of the BTRSS.
On the other hand, other experiment~\cite{Ekin,Alff,Iguchi,Wei,Sawa1,Aubin} did
not show the zero-field splitting. 
Thus the experimental results are still controversial at present.
There are some reasons which explain the contradiction in experiments.
One of them is the sample quality of the junction. 
Actually a group of experiment insists that the zero-field split would
be found when the sample quality is good enough and the measurement is done
in sufficiently low temperatures.
If this prediction is true, the impurity scattering might unify the splitting peaks. 
We try to check this prediction as follows. 

\begin{figure}
\begin{center}
\includegraphics[width=8.0cm]{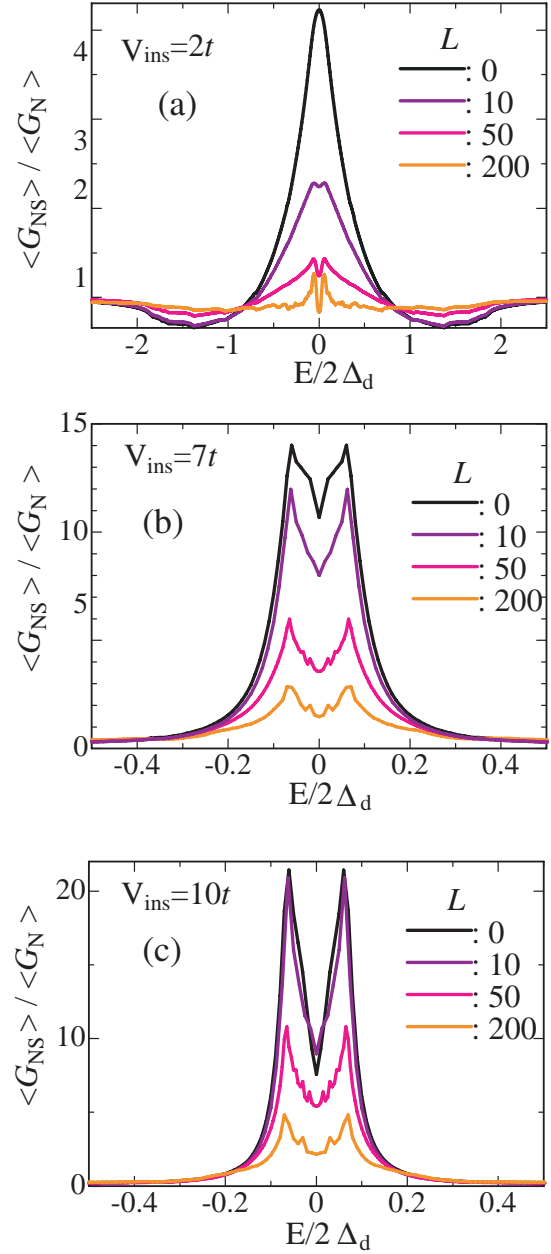}
\end{center}
\caption{ The conductance in DN/I/S is plotted as a function of the bias voltage,
where $M=32$, $\xi=10$ and $\Delta_s=0.2\Delta_d$. The potential of an insulating layer
are chosen to be $V_{ins}/t$ = 2, 7 and 10 in (a), (b) and (c), respectively. 
}
\label{fig:dis3}
\end{figure}

We introduce the impurity potential in normal metals of $d+is$-wave
junctions. In Fig.~\ref{fig:dis3}, the conductance is plotted 
as a function of the bias voltage for several $L$, where $V_{ins}/t$ = 2, 7 and 10 in 
(a), (b) and (c), respectively. 
The $s$-wave component is introduced 
only at the NS interface and $\xi$ in Eq.~(\ref{sufs}) is chosen to be $10$ lattice
constants. The amplitude of the 
$s$-wave pair potential is fixed at $\Delta_s=0.2\Delta_d$. 
When the transparency of the interface is high, 
the ZBCP in clean junctions does not split into two peaks even 
in the presence of the $s$-wave component as shown in (a). 
The splitting can be seen in the disordered junctions 
as shown in the results for finite $L$. 
Thus the impurity scattering in normal metal seems to assist 
the splitting of the ZBCP. 
At the same time, the amplitude of the conductance decreases with 
increasing $L$.
In highly transparent junctions, the calculated results indicate 
the opposite conclusion to the experimental prediction. 
When the transparency of the junction
becomes low, the ZBCP splits into two even in clean junctions as show in (b)
and (c). The amplitude of the conductance decreases with
the increase of the resistance of normal metals as well as those in (a). 
It is important that the degree of the splitting are not changed by the impurity scattering.
In low transparent junctions, our results are also contradict to the 
experimental prediction.
We conclude that the impurity scattering in low transparent junctions 
does not merge the splitting peaks into a single ZBCP. 

\section{Conclusions}
We numerically calculated the differential conductance in dirty normal metal/ 
insulator/ superconductor junctions by using the recursive Green function method.
In superconductors, we assumed $d+is$ pairing symmetry which breaks the time-reversal
symmetry of the junctions. In pure $s$-wave (pure $d$-wave) junctions, we 
confirmed the 
presence (absence) of the proximity effect, which agrees 
with the quasiclassical Green function theory.
In the case of the $d+is$ symmetry, the proximity effect recovers when the pair 
amplitude of the $s$-wave symmetry becomes dominant. 
In highly transparent clean junctions, we found the single ZBCP even in the absence of
the time-reversal symmetry \cite{Tanuma2001}. 
The ZBCP, however, splits into two peaks owing to the impurity scattering.
The $s$-wave component splits the zero-bias conductance peak into two peaks in
low transparent junctions even in the clean limit. 
The amplitudes of the splitting peaks
decreases with the increase of the impurity scattering in normal metals.
The peak position, however, remains unchanged even in the presence of the strong
impurity scattering. Thus we conclude that the impurity scattering does not
merge the two splitting peaks.

\acknowledgements 

This work was partially supported by the Core Research for
Evolutional Science and Technology (CREST) of the Japan Science and
Technology Corporation (JST). 
The computational aspect of this work has been performed at the 
facilities of the Superconputer Center, Institute of Solid State 
Physics, University of Tokyo and the Computer Center. 
J.I. acknowledges support by the NEDO 
international Joint Research project 
"Nano-scale Magnetoelectronics".

%
%


\begin{thebibliography}{99}
\bibitem{Tana1} Y.~Tanaka and S.~Kashiwaya: Phys. Rev. Lett. \textbf{74} (1995) 3451.

\bibitem{Alff}
L.~Alff, H.~Takashima, S.~Kashiwaya, N.~Terada, H.~Ihara,
Y.~Tanaka, M.~Koyanagi and K.~Kajimura: Phys. Rev. B \textbf{55} (1997) 14757.

\bibitem{Wei}
J.~Y.~T.~Wei, N.-C.~Yeh, D.~F.~Garrigus and M.~Strasik:
Phys. Rev. Lett. \textbf{81} (1998) 2542.

\bibitem{Wang}
W.~Wang, M.~Yamazaki, K.~Lee and I.~Iguchi: Phys. Rev. B \textbf{60} (1999) 4272.

\bibitem{Iguchi} I.~Iguchi, W.~Wang, M.~Yamazaki, Y.~Tanaka, and S.~Kashiwaya:
Phys. Rev. B \textbf{62} (2000) R6131.

\bibitem{Kashi2000} S.~Kashiwaya and Y.~Tanaka: Rep. Prog. Phys. \textbf{63} (2000) 1641.

\bibitem{Kashi95} S.~Kashiwaya, Y.~Tanaka, M.~Koyanagi, H.~Takashima and K.~Kajimura:
Phys. Rev. B \textbf{51} (1995) 1350.

\bibitem{Kashi96} S.~Kashiwaya, Y.~Tanaka, M.~Koyanagi and K.~Kajimura: Phys. Rev. B \textbf{53} (1996) 2667.

\bibitem{Sawa1} A. Sawa, S. Kashiwaya, H. Obara, H. Yamasaki, M. Koyanagi, Y. Tanaka 
and N. Yoshida: Physica C \textbf{339} (2000) 107.  

\bibitem{Sawa2} H. Kashiwaya, A. Sawa, S. Kashiwaya, H. Yamazaki, M. Koyanagi, 
I. Kurosawa, Y. Tanaka, and I. Iguchi:  Physica C \textbf{357}-\textbf{360} (2001) 1610.

\bibitem{Hu} C.~R.~Hu: Phys. Rev. Lett. \textbf{72} (1994) 1526.

\bibitem{matsumoto}
M. Matsumoto and H. Shiba, J. Phys. Soc. Jpn. 
\textbf{64} (1995) 1703. 

\bibitem{nagato} Y. Nagato and K. Nagai, Phys. Rev. B 
\textbf{51} (1995) 16254.   

\bibitem{ohhashi}
Y. Ohashi, J. Phys. Soc. Jpn. \textbf{65} (1996) 823; 
Y. Ohashi and S. Takada,  J. Phys. Soc. Jpn. \textbf{65} (1996) 246.  

\bibitem{sign1} Y.~Tanuma, Y.~Tanaka, M.~Yamashiro and S.~Kashiwaya
: Phys. Rev. B \textbf{57}  (1998) 7997; Y.~Tanuma, Y.~Tanaka, K.~Kuroki and S.~Kashiwaya: 
Phys. Rev. B \textbf{64} (2001) 214510; 
Y.~Tanuma, K.~Kuroki Y.~Tanaka, R.~Arita, S.~Kashiwaya and H.~Aoki: Phys. Rev. B \textbf{66} (2002) 094507; Y.~Tanuma, K.~Kuroki Y.~Tanaka, and  S.~Kashiwaya: 
Phys. Rev. B \textbf{66} (2002) 174502. 


\bibitem{sign2}
M. Yamashiro, Y. Tanaka and S. Kashiwaya: Phys. Rev. B \textbf{56}  (1997) 7847; 
M. Yamashiro, Y. Tanaka Y. Tanuma and S. Kashiwaya: 
J. Phys. Soc. Jpn. \textbf{68} (1999) 2019; 
M. Yamashiro, Y. Tanaka N. Yoshida 
and S. Kashiwaya: J. Phys. Soc. Jpn. \textbf{67} (1998) 3224. 

\bibitem{sign3}
Y. Tanaka, H. Tsuchiura, Y. Tanuma and S. Kashiwaya: 
J. Phys. Soc. Jpn. \textbf{71} (2002) 271; 
Y. Tanaka, Y. Tanuma  K. Kuroki and S. Kashiwaya: 
J. Phys. Soc. Jpn. \textbf{71} (2002) 2102. 

\bibitem{Zhu2}
H.X. Tang, Jian-Xin Zhu and Z.D. Wang: Phys. Rev. B {\bf 54} 
(1996) 12509; Jian-Xin Zhu H. X. Tang and Z.D. Wang: Phys. Rev. B {\bf 54} 
(1996) 7354.  





\bibitem{tanaka6}
Y. Tanaka: Phys. Rev. Lett. {\bf 72} (1994) 3871;  
Y.~Tanaka and S.~Kashiwaya: Phys. Rev. B \textbf{53} (1996) 11957; 
Phys. Rev. B \textbf{56}  (1997) 892;  
Phys. Rev. B \textbf{58}  (1998) R2948.  

\bibitem{barash} Y.~S.~Barash, H.~Burkhardt and D.~Rainer,
Phys. Rev. Lett. \textbf{77} (1996) 4070.

\bibitem{t61}
Y.~Tanaka and S.~Kashiwaya: 
J. Phys. Soc. Jpn.  \textbf{68}  (1999) 3485;  
J. Phys. Soc. Jpn.  \textbf{69}  (2000) 1152.  

\bibitem{Kusakabe}
Y. Tanaka, T. Hirai, K. Kusakabe and S. Kashiwaya: 
Phys. Rev. B \textbf{60}  (1999) 6308;
T. Hirai, K. Kusakabe and Y. Tanaka
Physica C \textbf{336} (2000)  107; 
K. Kusakabe and Y. Tanaka; Physica C 
\textbf{367} (2002) 123; 
K. Kusakabe and Y. Tanaka; J. Phys. Chem. Solids
\textbf{63} (2002) 1511.  


\bibitem{asano01-3} Y.~Asano: Phys. Rev. B \textbf{64} (2001) 224515.

\bibitem{asano02-3} Y.~Asano and K.~Katabuchi: J. Phys. Soc. Jpn. \textbf{71} (2002) 1974.

\bibitem{asano03-1} Y.~Asano, Y.~Tanaka, M.~Sigrist and S.~Kashiwaya: cond-mat/0212353.


\bibitem{Zhu} J-X. Zhu, B. Friedman, and C. S. Ting: 
Phys. Rev. B \textbf{59} (1999) 9558.
\bibitem{Kashi2}
S. Kashiwaya \textit{et al.}: Phys. Rev. B \textbf{60} (1999) 3527.
\bibitem{Zutic}
I. Zutic and O. T. Valls: Phys. Rev. B \textbf{60} (1999) 6320.
\bibitem{Yoshida} N. Yoshida, Y. Tanaka, J. Inoue, and S. Kashiwaya: 
J. Phys. Soc. Jpn. \textbf{68} (1999) 1071. 
\bibitem{Hirai} T. Hirai, N. Yoshida, Y. Tanaka, J. Inoue and S. Kashiwaya: 
J. Phys. Soc. Jpn. \textbf{70} (2001) 1885. 
\bibitem{Yoshida2002} N. Yoshida, H. Itoh, T. Hirai, Y. Tanaka, J. Inoue and S. Kashiwaya: 
Phsica C \textbf{367} (2002) 135.


\bibitem{fogelstrom} M.~Fogelstr\"{o}m, D.~Rainer, and J.~A.~Sauls:
Phys. Rev. Lett. \textbf{79} (1997) 281;
D.~Rainer, H.~Burkhardt, M.~Fogelstr\"{o}m, and J.~A.~Sauls:
J. Phys. Chem. Solids \textbf{59} (1998) 2040.

\bibitem{YT021} Y. Tanaka, H. Tsuchiura, Y. Tanuma and S. Kashiwaya:  
J. Phys. Soc. Jpn.  \textbf{71}  (2002) 271.

\bibitem{YT022}
Y. Tanaka, H. Itoh, H. Tsuchiura, Y. Tanuma, J. Inoue, 
and S. Kashiwya: J. Phys. Soc.  Jpn. \textbf{71} (2002) 2005. 

\bibitem{YT023}
Y. Tanaka, Y. Tanuma, K. Kuroki and S. Kashiwaya:
J. Phys. Soc.  Jpn. \textbf{71}  (2002) 2102. 

\bibitem{Kashiwaya95}
S. Kashiwaya, Y. Tanaka, M. Koyanagi and K. Kjimura, 
J. Phys. Chem. Solids \textbf{56} (1995) 1721. 
%
\bibitem{TJ1}
Y. Tanuma, Y. Tanaka, M. Ogata and S. Kashiwaya:
J. Phys. Soc. Jpn., \textbf{67} (1998) 1118. 
%
\bibitem{TJ2}
Y.~Tanuma, Y.~Tanaka, M. Ogata and S.~Kashiwaya: Phys. Rev. B \textbf{60} (1999) 9817.


\bibitem{Tanuma2001} Y. Tanuma, Y. Tanaka, and S. Kashiwaya:
Rhys. Rev. B \textbf{64} (2001) 214519.

\bibitem{Zhu1}
Jian-Xin Zhu, B. Friedman and C.S. Ting: Phys. Rev. B {\bf 59} (1999) 3353; 
Jian-Xin Zhu and C.S. Ting: Phys. Rev. B {\bf 60} (1999) R3739; 
Jian-Xin Zhu  and C.S. Ting: Phys. Rev. B {\bf 57} (1998) 3038.


\bibitem{matsumoto2} M.~Matsumoto and H.~Shiba: 
J. Phys. Soc. Jpn. \textbf{64} (1995) 3384; 
J. Phys. Soc. Jpn. \textbf{64} (1995) 4867.

\bibitem{laughlin} R.~B.~Laughlin: Phys. Rev. Lett. \textbf{80} (1998) 5188.

\bibitem{tanaka3} Y.~Tanaka, Y.~Tanuma, and S.~Kashiwaya:
Phys. Rev. B \textbf{64} (2001) 054510.

\bibitem{tanaka4}
Y. Tanaka, T. Asai, N. Yoshida, J. Inoue and S. Kashiwaya:  
Phys. Rev. B  \textbf{61} (2000)  R11902. 

\bibitem{covington} M.~Covington, M.~Aprili, E.~Paraoanu, L.~H.~Greene,
F.~Xu, J.~Zhu, and C.~A.~Mirkin:
Phys. Rev. Lett. \textbf{79} (1997) 277; 
M. Aprili, E. Badica and L. H. Greene 
Phys. Rev. Lett. \textbf{83} (1999) 4630.  

\bibitem{biswas} A.~Biswas, P.~Fournier, M.~M.~Qazilbash, V.~N.~Smolyaninova, H.~Balci, 
and R.~L.~Greene: Phys. Rev. Lett. \textbf{88} (2002) 207004.

\bibitem{dagan} Y.~Dagan and G.~Deutscher:
Phys. Rev. Lett. \textbf{87} (2001) 177004.

\bibitem{sharoni} A.~Sharoni, O.~Millo, A.~Kohen, Y.~Dagan, R.~Beck, G.~Deutscher, and G.~Koren:
Phys. Rev. B \textbf{65} (2002) 134526.

\bibitem{Ekin}
J. W. Ekin, Y. Xu, S. Mao, T. Venkatesan, D. W. Face, M. Eddy, 
and S. A. Wolf: Phys. Rev. B \textbf{56} (1997) 13746. 


\bibitem{Aubin} 
H. Aubin, L. H. Greene, Sha Jian and D. G. Hinks:
Phys. Rev. Lett. \textbf{89} (2002) 177001.


\bibitem{Neils}
W. K. Neils and D. J. Van Harlingen: 
Phys. Rev. Lett. \textbf{88} (2002) 047001. 

\bibitem{asano02-1} Y.~Asano and Y.~Tanaka: Phys. Rev. B \textbf{65} (2002) 064522. 

\bibitem{asano03-2} Y.~Asano, Y.~Tanaka and S.~Kashiwaya: cond-mat/0302287.




\bibitem{eilenberger} G.~Eilenberger: Z. Phys. \textbf{214} (1968) 195.

\bibitem{larkin} A.~I.~Larkin and Yu.~N.~Ovchinikov: Eksp. Teor. Fiz.
\textbf{55} (1968) 2262.[Sov. Phys. JETP \textbf{28} (1968) 1200.]

\bibitem{zaitsev} A.~V.~Zaitsev: Zh. Eksp. Teor. Fiz. \textbf{86} (1984) 1742.
[Sov. Phys. JETP \textbf{59} (1984) 1015.]


\bibitem{shelankov} A.~L.~Schelankov: J. Low. Tem. Phys. 
\textbf{60} (1985) 29.

\bibitem{bruder} C.~Bruder: Phys. Rev. B \textbf{41} (1990) 4017.

\bibitem{barash2} Y.~S.~Barash, A.~A.~Svidzinsky and H.~Burkhardt:
Phys. Rev. B \textbf{55} (1997) 15282.

\bibitem{golubov} A.~A.~Golubov, M.~Y.~Kupriyanov: Pis'ma Zh. Eksp. Teor. fiz 
\textbf{69} (1999) 242.[ Sov. Phys. JETP Lett. \textbf{69} (1999) 262.];
\textbf{67} (1998) 478.[ Sov. Phys. JETP Lett. \textbf{67} (1998) 501.]

\bibitem{poenicke} A.~Poenicke, Yu.~S.~Barash, C.~Bruder, and V.~Istyukov:
Phys. Rev. B \textbf{59} (1999) 7102.

\bibitem{yamada} K.~Yamada, Y.~Nagato, S.~Higashitani and K.~Nagai:
J. Phys. Soc. Jpn. \textbf{65} (1996) 1540.


\bibitem{luck} T.~L\"{u}ck, U.~Eckern, and A.~Shelankov:
Phys. Rev. B \textbf{63} (2002) 064510.
%
\bibitem{Greene}  Discussion at M$^{2}$S  2002.
\bibitem{Beenakker1} C. W. J. Beenakker: Rev. Mod. Phys. \textbf{69} (1997) 731.

\bibitem{Lambert} C. J. Lambert: J. Phys. Condens. Matter \textbf{3} (1991) 6579.

\bibitem{Takane} Y. Takane and H. Ebisawa: J. Phys. Soc. Jpn. \textbf{61}
(1992) 2858.

\bibitem{Lesovik} G. B. Lesovik, A. L. Fauchere, and G. Blatter: Phys. Rev. B
\textbf{55} (1997) 3146.


\bibitem{Beenakker2} C. W. J. Beenakker: Phys. Rev. B \textbf{46} (1992) 12841.

\bibitem{Beenakker3} C. W. J. Beenakker, B. Rejaei and J. A. Melsen: Phys. Rev.
Lett. \textbf{72} (1994) 2470.


\bibitem{Volkov} A. F. Volkov, A. V. Zaitsev and T. M. Klapwijk: Physica C \textbf{210}
(1993) 21.


\bibitem{Yip} S. Yip: Phys. Rev. B \textbf{52} (1995) 3087.

\bibitem{Golubov2} A. A. Golubov, F. K. Wilhelm, and A. D. Zaikin: Phys. Rev. B
\textbf{55} (1997) 1123.

\bibitem{Takayanagi} A. F. Volkov and H. Takayanagi: Phys. Rev. B \textbf{56}
(1997) 11184.



%

\bibitem{Nazarov1} Yu. V. Nazarov: Phys. Rev. Lett. \textbf{73} (1994) 1420.

\bibitem{Nazarov2} Yu. V. Nazarov: Superlatt. Microstruct. \textbf{25} (1999) 1221;
cond-mat/9811155.


\bibitem{Golubov2003} Y. Tanaka, A. Golubov and S. Kashiwaya: 
Phys. Rev. B  2003 in press. 




\bibitem{Tanaka2002} Y. Tanaka, Yu. V. Nazarov and S. Kashiwaya: 
cond-mat/0208009. 

\bibitem{asano01-1} Y.~Asano: Phys. Rev. B \textbf{63} (2001) 052512.

\bibitem{asano01-2} Y.~Asano: Phys. Rev. B \textbf{64} (2001) 014511.

\bibitem{asano02-2} Y.~Asano: J Phys. Soc. Jpn. \textbf{71} (2002) 905.

\bibitem{Itoh1} H. Itoh, Y. Tanaka, J. Inoue, and S. Kashiwaya: 
Physica C \textbf{367} (2002) 99; 
N. Yoshida, Y. Asano, H. Itoh, Y. Tanaka, J. Inoue and S. Kashiwaya 
J. Phys. Soc. Jpn. \textbf{72} No. 4 (2003)

\bibitem{Itoh2}
H. Itoh,  N. Kiraura, Y. Yoshida, Y. Tanaka, J. Inoue, 
Y. Asano and S. Kashiwaya: "\textit{Toward the Controllable Qunatum States}", 
Eds. H. Takayanagi and J. Nitta, pp. 173, World Scientific Publishing (2003). 
 

%


\bibitem{lee} P. A. Lee and D. S. Fisher: Phys. Rev. Lett. \textbf{47} (1981) 882.

\bibitem{kubo} R. Kubo: J. Phys. Soc. Jpn. \textbf{12} (1957) 570. 

\bibitem{blonder} G.~E.~Blonder, M.~Tinkham and T.~M.~Klapwijk:
 Phys. Rev. B \textbf{25} (1982) 4515.

\bibitem{takane92} Y. Takane and H. Ebisawa: J. Phys. Soc. Jpn. \textbf{61} (1992) 1685. 

\bibitem{asano02-4} Y.~Asano: Phys. Rev. B \textbf{66} (2002) 174506.


\end{thebibliography}
\end{document}